\documentclass[a4paper,11pt]{article}

\usepackage{jinstpub} 
\usepackage{verbatim}
\usepackage[compact]{titlesec}
\usepackage{xurl}
\hypersetup{breaklinks=true}

\usepackage{lineno}
\usepackage[version=4]{mhchem}
\usepackage{mathtools}

\title{\boldmath Primary scintillation yields induced by $\alpha$ particles in gas mixtures of Argon/CF$_4$ at 9.5~bar}

\author[a]{P. Amedo,}
\author[a]{S. Leardini,}
\author[a]{A. Saá-Hernández,}
\author[a]{D. González-Díaz}

\affiliation[a]{Instituto Gallego de Física de Altas Energías, Univ. de Santiago de Compostela, Campus sur, Rúa Xosé María Suárez Núñez, s/n, Santiago de Compostela, E-15782, Spain}

\emailAdd{diego.gonzalez.diaz@usc.es}

\abstract{We report time and band -resolved scintillation  from $\alpha$-particles in Ar/CF$_4$ mixtures at a pressure of 9.5~bar. Our results show that \%-level addition of CF$_4$ enables strong wavelength-shifted scintillation in the visible range, with yields at the level of 1000~ph/MeV, scintillation decay times of 9-25~ns and formation times well below 10~ns. Such a performance is a priori sufficient for accurate time-tagging of MeV-particles, without the need to resort to a pure noble gas, thus opening an appealing technological path for next-generation time projection chambers.}


\makeatletter
\gdef\@fpheader{}
\makeatother

\begin{document}
\maketitle
\flushbottom

\section{Introduction}

Tetrafluoromethane (CF$_4$) is a frequent guest to particle physics instrumentation \cite{Battat2014, ARIADNE_CF4, Fraga:2002uc, Takahashi2011, RICH_LHCb,Fraga_GEM}, one of the reasons being its strong VUV-visible scintillation strength of about $1000$-$3000$~ph/MeV under ionizing radiation \cite{Pansky1995, Morozov2010, Azmoun:2010zz, Lehaut2015}, around a factor of 10 shy of that in noble gases \cite{Saito_Ws_NobleGases_2002}. While this performance may seem modest, CF$_4$ enjoys a number of other remarkable features such as scintillation time constants below $15$~ns \cite{Morozov2011} (considerably shorter than that of the triplet emission from noble gases \cite{xenon_paper, Santorelli2021}), high transparency down to VUV \cite{Spectral_Atlas_2013}, and easiness of light collection and detection (most of its scintillation takes place in the range 200-800~nm \cite{Amedo2023}). It possesses a strong electron-cooling power \cite{Christophorou2004} (hence, low diffusion) and, in admixtures with argon, it enables optical tracking down to minimum ionizing particles, through avalanche-induced scintillation \cite{Amedo_tracks_2024}. Therefore, when considering detection systems that require the simultaneous detection of ionization and scintillation, and good track fidelity (see, e.g., \cite{DiegoReview2018} and references therein), CF$_4$ is arguably on a league of its own.

A subtler motivation to study systems based on CF$_4$ is the intriguing possibility that, in a way analogous to how `Penning mixtures' enhance ionization \cite{Sain2010_Penning, Sahin2016_Penning}, a suitable `wavelength-shifting mixture' can potentially render scintillation strengths larger than those in the pure gas \cite{Forster_1959}, shifting it to regions where it is easier to detect:
\begin{eqnarray}
    A^{**} + B \rightarrow & A + B^+ + e^- & ~~~~\textnormal{Penning transfer}\\
    A^{**} + C \rightarrow & A + C^* \rightarrow A + C + h\nu & ~~~~\textnormal{gaseous wavelength-shifting} \label{wlsE}
\end{eqnarray}
(here $^{**}$ represents, generically, an excited state up to the ionization continuum and $A$ and $B, C$ are atomic or molecular species). Wavelength-shifting, when performed directly in gas phase instead of through solid converters (e.g., TPB\cite{Benson2018}, PEN\cite{Kuzniak_2021_FATGEMS}...) has other potential advantages: i) it can be expected to be more cost-effective and simpler to implement, and applicable even in trace-amount concentrations (e.g., \cite{DUNE_Xe_Doping_2024}) and ii) can provide a faster time response, given that it is not needed to wait for the $A^{**}$ state to emit (e.g. \cite{Segreto_TPB_LAr_2015}). 

Compared to the body of knowledge amassed in regard to the CF$_4$ scintillation properties, the behaviour of CF$_4$ as a potential gaseous wavelength-shifter is much less studied, which is surprising since several experiments based on time projection chambers (TPCs) use or aim at making use of CF$_4$-based mixtures \cite{CYGNUS,Migdal_2023,NDGAR_Snowmass2021,Cortesi2023}. In fact, and to the best of the authors knowledge, no systematic studies of the primary scintillation yields seem to exist in literature, when CF$_4$ is used as a wavelength-shifter instead of as the main gas. A spectral analysis performed by our group in \cite{Amedo2023} allowed to address the strength of the transfer process \ref{wlsE} in the primary scintillation of Ar/CF$_4$ relative to pure Ar and CF$_4$ phases, for X-ray irradiation. That work shows for instance that 2\%-addition of CF$_4$ to argon is sufficient to match the amount of visible scintillation in pure CF$_4$, with the yields increasing for higher concentrations up to 10\%CF$_4$, at least. Based on the yields reported for pure CF$_4$, this observation would seemingly argue for scintillation strengths in the range of few 1000's~ph/MeV for Ar/CF$_4$ mixtures too. The differences reported in \cite{Amedo2023}in the spectra of emission for $\alpha$ particles and X-rays (up to a factor $\times 3$), however, and the absence of absolute-yield determinations for the latter study, preclude a direct estimate.

Our main motivation in this work is to assess the effectiveness of reactions of the type \ref{wlsE} in Ar/CF$_4$ admixtures, through a direct measurement of the scintillation yields and time profiles. We target specifically future neutrino experiments performed with pressurized argon TPCs (as presently intended by the DUNE collaboration \cite{NDGAR_Snowmass2021,DUNE_Phase_II}, or under consideration for the upgrade of COHERENT \cite{COHERENT_Snowmass_2021}), for which we explored a range of CF$_4$ concentrations from 0.1\% up to 10\%  (per volume), at a pressure around 10~bar. According to simulations performed in \cite{Ar_CF4_Timing_2024}, an optical response in the range of 1000 ph/MeV with time constants of few 10's of ns would enable time tagging ($T_0$) with ns-level accuracy, in future neutrino detectors such as the high pressure TPC of DUNE's ND-GAr.

This paper comprises two main sections: section II recalls our experimental setup and procedures in a succinct way, and section III systematizes the main experimental results of our study. We end with our conclusions in section IV.

\section{Experimental setup and data analysis}

Measurements were performed following a small adaptation of the device employed in \cite{xenon_paper}: a mini time projection chamber (TPC) instrumented with a multi-wire readout, assembled in an all-CF/VCR gas system within a CF100 stainless-steel vessel, equipped with a recirculation system and purity-control. Compared to our earlier mini-TPC built and characterized thoroughly in Xe, the present one has been optimized for lower scintillation levels and therefore the drift-region was reduced to 0.7~cm. This brings the radioactive source closer to the optical system and eliminates the need for field-shapers. As in previous measurements, the chamber achieved a vacuum level of $10^{-3}$ mbar. The filling was done directly from the Ar and CF$_4$ bottles (purity 5 and 4.7, respectively) up to the desired pressure. Gas was recirculated to assure homogeneous mixing. It was verified during the campaign that the N$_2$ concentration levels were below 0.1\% at all times, the sensitivity being limited by the background of the residual gas analyzer. It has been shown in earlier works (\cite{Margato}) that CF$_4$ scintillation is robust up to N$_2$ concentrations as high as 4\% (factor $\times 2$ drop), so we anticipate no purity issues in the conditions of this work.

The anode region was instrumented with thin wires to ensure good optical transparency towards the photosensor plane placed behind it. All four planes' wires were made of a 99.95\% purity tungsten core, plated with 99.99\% purity gold. Anode wires had a diameter of 20 $\pm$ 2 $\mu$m and the rest of planes were assembled with 80~$\pm$~8~$\mu$m -diameter ones. Anode and cathode wires have a pitch of 2.5~mm (shifted by half a pitch for each alternating plane), while the gating grid has a pitch of 1.25~mm. A metallic disk with an $^{241}$Am deposit was housed in a circular groove at the middle of the TPC-cathode, leaving it flush with its surface and thus ensuring a minimal distortion of the electric field. 

To increase sensitivity in the visible region compared to our earlier setup in \cite{xenon_paper}, four pressure-resistant photomultipliers (PMs), model R7378 ($\times 3$) and 5070 ($\times 1$) from Hamamatsu, were assembled in a teflon frame and placed at close distance. A coincidence between two of the R7378 PMs was used to set the trigger, and data from the other two was used in the analysis. The trigger was set generally at the level of few photoelectrons. As the length of an $\alpha$ track is below 5~mm for argon at around 9.5~bar (much smaller than the dimensions of the optical system) events could be considered point-like for practical purposes and no trigger bias is anticipated. This was confirmed by triggering on the secondary scintillation signal and looking at the primary scintillation yields, that provided results compatible with the analysis presented here.

\begin{figure}[h!!!]
    \centering
    \includegraphics[scale=0.6]{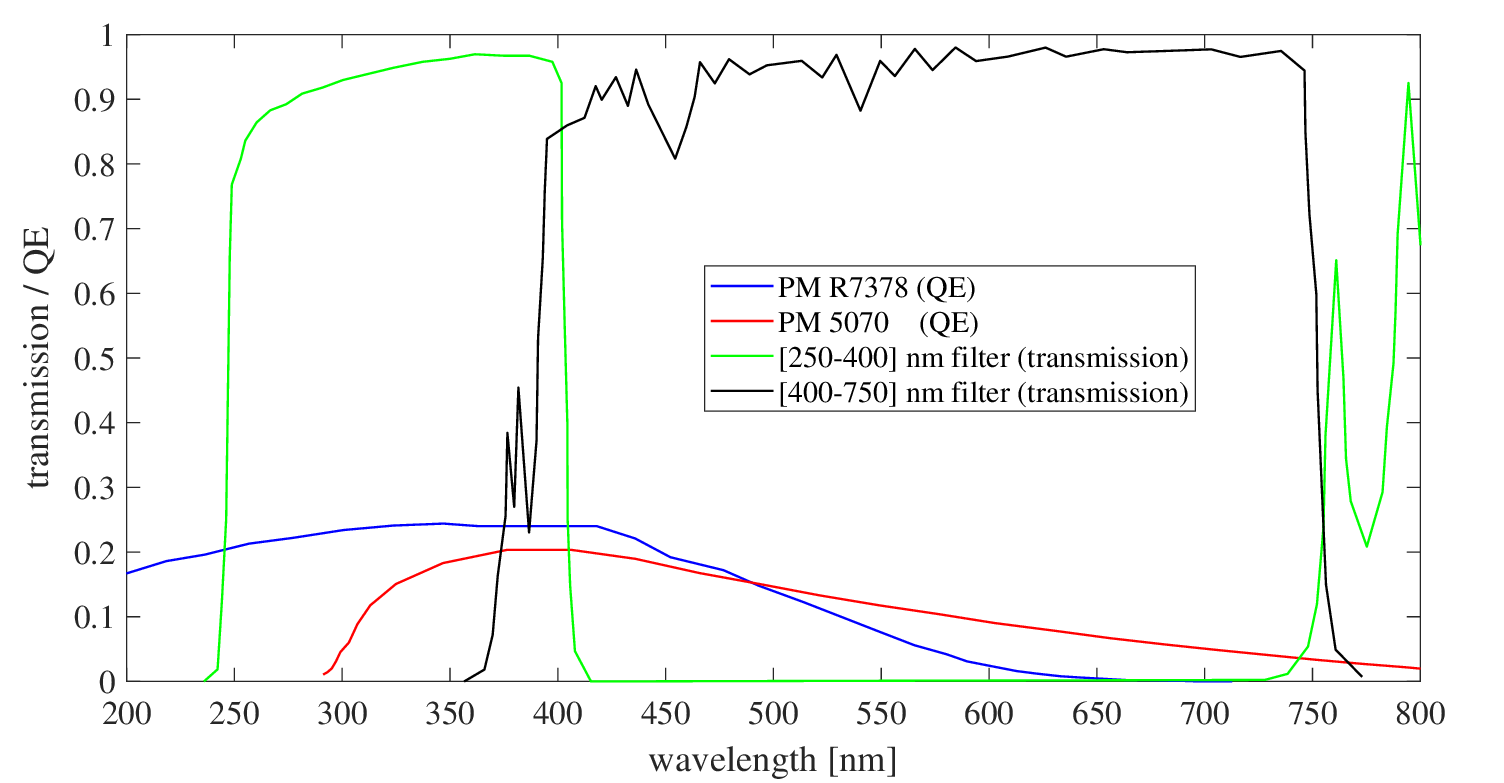}
    \caption{Characteristics of the main optical elements used for the TPC readout. Three R7378 PMs from Hamamatsu ($QE$ in blue) are used, one of them coupled to a UV-filter (green line). Measurements in the visible region are performed with a 5070 PM also from Hamamatsu, coupled to a visible-filter (black line).}
\label{fig:Transm}
\end{figure}

PM model R7378 was used, coupled to a band-pass UV-filter from Asahi ([250-400]~nm), in order to assess the UV emission. The tandem covers most of the UV band of CF$_4$ and, with a nearly flat response at around 20-22\%, it makes measurements in this range little dependent on the spectral content (Fig. \ref{fig:Transm}, green and blue data). For the visible range, PM model 5070 coupled to a band-pass visible-filter ([400-750]~nm), also from Asahi, was used (Fig. \ref{fig:Transm}, black and red data). Quantum efficiencies were weighted with the spectra measured in \cite{Amedo2023} at 5~bar, yielding values of $QE_{uv} = 21\%$ and $QE_{vis} = 8.0\%$. Shape-variations of the spectra, in the range of CF$_4$ concentrations explored here, caused relative variations of less than 0.2\% and 2\% in the estimated $QE$'s, respectively.

\begin{figure}[h!!!]
    \centering
    \includegraphics[scale=0.65]{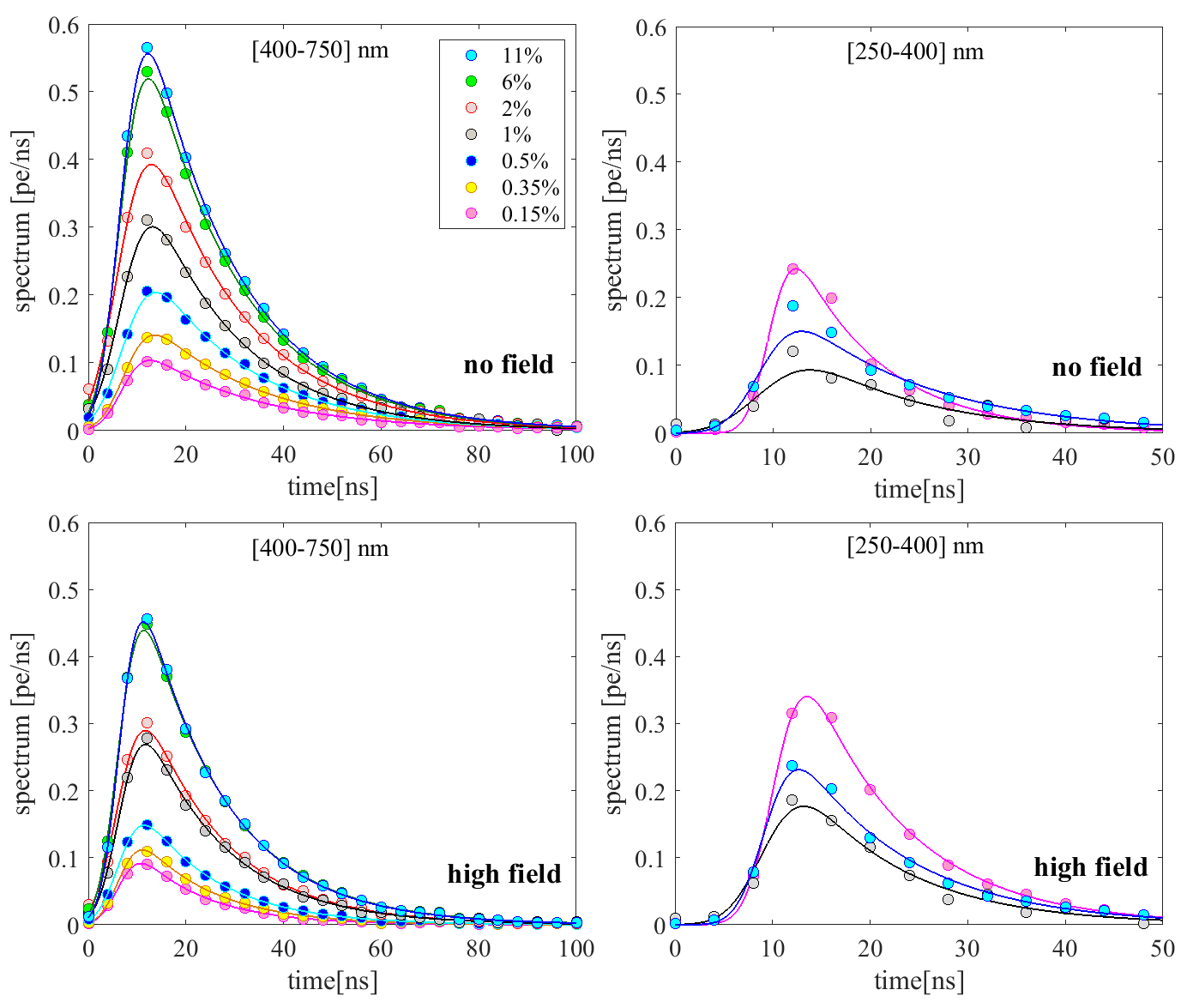}
    \caption{Average waveforms obtained upon excitation with a $^{241}$Am source, for different CF$_4$ concentrations (in photoelectrons per ns). Data is shown as circles. A fit to the convolution of a Gaussian time-response (from the PM) and an exponential decay (from the gas) is super-imposed. Left figures show the emission in the visible range and right figures the one in the UV range. Top row presents the results for no electric field and the bottom row for a pressure-reduced electric field around 100~V/cm/bar. The operating pressure is 9.5~bar, varying within 0.2~bar (maximum) from run to run. For the UV band there is no lack of systematic behaviour (for an explanation see later in text), so only three representative concentrations are shown in order to avoid overcrowding.}
\label{fig:wvfs}
\end{figure}

Data acquisition was performed with a CAEN DT5725 board of 14~bits, 125~MHz bandwidth and a sampling time of 4~ns. The acquisition window was set to 1~$\mu$s for S1 data and to 2~$\mu$s for S1-S2 combined data.\footnote{We resort to a widely-used notation these days: S1 refers to the primary scintillation signal and S2 to the secondary scintillation one.} The latter runs were used to assess the overall consistency of the measurements, providing an independent study of the drift velocity and S1 yields. The setup was initially intended to offer a glimpse into the optical gains achievable in a MWPC system, however the obtained values of around 10 ph/e are probably limited due to the short distance between the MWPC plane and the Bragg peak of the alpha-track (down to few mm). Fig. \ref{fig:wvfs} presents (circles) an example of the average waveforms taken, in photoelectrons/ns, for conditions of no field and high field (around 100~V/cm/bar), at a pressure of 9.5~bar.

Data analysis was performed with Matlab scripts, including basic functions for pulse-shape analysis such as baseline corrections and waveform time-alignment to correct for offsets. Characteristic waveform parameters (signal rise-time, fall-time, amplitude, event barycenter, integrated charge) were then retrieved. The influence of applying cuts on those variables was studied, but their impact on the results found to be negligible. Calibration from PM charge to photoelectrons was performed with a pulsed LED near the single-photon level (as described in \cite{xenon_paper}), and the influence of the PM response function on the time spectra modeled as:
\begin{equation}
\mathcal{F}(t) = (A e^{-t/\tau_{UV(vis)}}) \circledast G(\sigma_, t_o) \label{ExpFunc}
\end{equation}
Here $G(\sigma_, t_o)$ is the PM time-response function under $\delta$-excitation (assumed to be a Gaussian of width $\sigma$ and offset $t_0$) and the symbol $\circledast$ denotes a convolution. With exception of UV data at low fields, the data was well fitted to the above `ansatz' (Fig. \ref{fig:wvfs}, lines). Despite a second time constant would make the fit better in this latter case, the difference was deemed too small and eq. \ref{ExpFunc} was preserved for the sake of simplicity.

\section{Main results}

\begin{figure}[h!!!]
    \centering
    \includegraphics[scale=0.53]{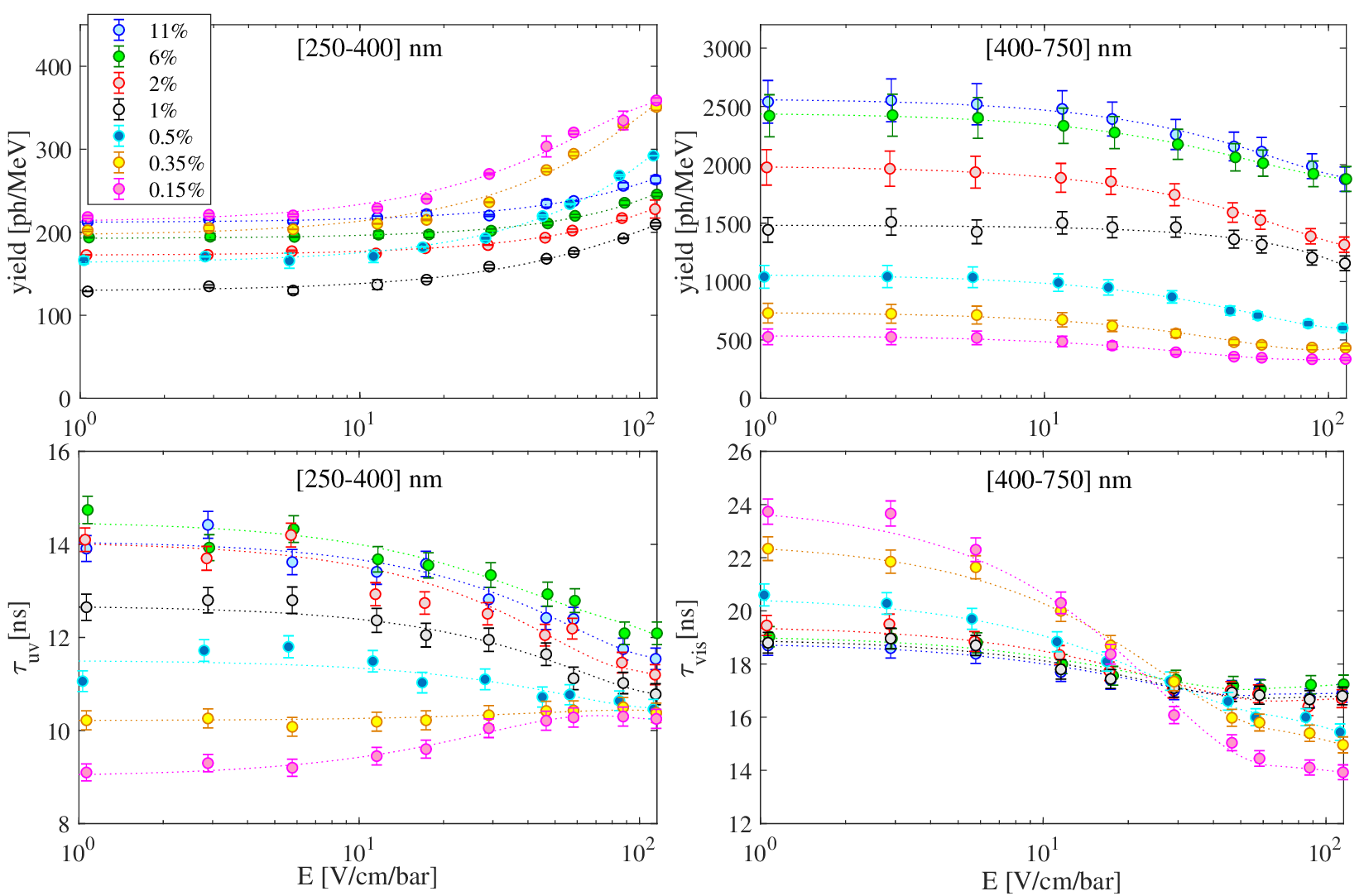}
    \caption{Systematization of Ar/CF$_4$ scintillation, based on the two parameters characterizing the emission: absolute yield and decay time. The top row shows the yields in the UV (left) and visible (right) bands. The bottom row shows the time constants. Following earlier works, the dependence with the electric field might be attributed to recombination, transferring energy from the CF$_4^{+,*}$ (UV) band to the CF$_3^{*}$ (visible) one. Dashed lines are splines to guide the eye.}
\label{fig:AnaParam}
\end{figure}

Data was taken as a function of the electric field, and systematized from the fits to eq.  \ref{ExpFunc} in terms of scintillation yields and time constants, as shown in Fig. \ref{fig:AnaParam}. The geometrical efficiency (probability of a photon landing on the PM photocathode) was estimated from Geant4 to be 0.0146 using the simulation framework developed in \cite{xenon_paper}. The overall uncertainty of the absolute normalization was estimated at 21\% (15\% from the single-photon calibration and 15\% from the estimate of the geometrical efficiency). Moreover, a dedicated measurement was performed under pure xenon, by resorting to data from one of the unfiltered R7378 PMs (a single-PM trigger was sufficient in those conditions). The $2^{\textnormal{nd}}$-continuum yields obtained in that reference run were compatible with the ones obtained in \cite{xenon_paper}, within the given uncertainties.

Data in Fig. \ref{fig:AnaParam} displays a rich phenomenology, with the anti-correlation between UV and visible yields at high electric fields resembling early observations for pure CF$_4$ in \cite{Morozov2010}, above 1~bar. In that work, such anticorrelation was naturally attributed to the scintillation stemming from charge recombination, moving energy from the CF$^{+,*}_4$ (UV) band into the CF$_3^*$ (visible) one. Our data is compatible with that hypothesis: one can observe for instance that for the lowest-quenched mixtures, where electron diffusion is higher (e.g., for 0.15-0.35\% CF$_4$ concentration), the yields in the visible region reach a plateau at the highest fields. For higher concentrations (lower diffusion) the maximum field applied in present conditions is not sufficient to reach the plateau of scintillation, suggesting that some recombination is still present and the effect is in fact stronger in that situation.

Time constants evolve greatly with electric field, too. For high field values, decay times in the visible band fit well to the range of $\tau=10$-$15$~ns reported in \cite{Morozov2011} for the CF$_3^*$ scintillation in case of pure CF$_4$, at mildly-high pressure (3~bar). The elongation at low fields reported in that work is also reproduced here. The UV case is more complicated due to the existence of an interplay between the Ar $3^{\textnormal{rd}}$ continuum and CF$_4^{+,*}$ scintillation \cite{Amedo2023}. In fact, the observed time constants, despite being generally shorter than those in the visible, are much longer than the ones reported in \cite{Morozov2011} in the range [260-370]~nm ($\tau=5$~ns). In both UV and visible bands, the increase of the time constants at low electric fields can be naturally interpreted due to additional recombination dynamics of the CF$_4^{+,*}$ ions and associated reactions. For the lowest concentrations, however, in case of the UV band (yellow and magenta circles in Fig. \ref{fig:AnaParam} bottom-left), it is likely that the contribution from the $3^{\textnormal{rd}}$ continuum plays a role too. A significant portion of that emission leaks into the [250-400]~nm region \cite{Amedo2023}, with a time constant around 5~ns \cite{Santorelli2021}. Last, although the formation time of the scintillation precursors, be it through transfer reactions or recombination, is difficult to access in present conditions and specially given the prescription used in the fit, a safe upper bound of 10~ns (from 10\% to 90\% of the signal value) can be estimated from the waveforms of Fig. \ref{fig:wvfs}.

\begin{figure}[h!!!]
    \centering
    \includegraphics[scale=0.55]{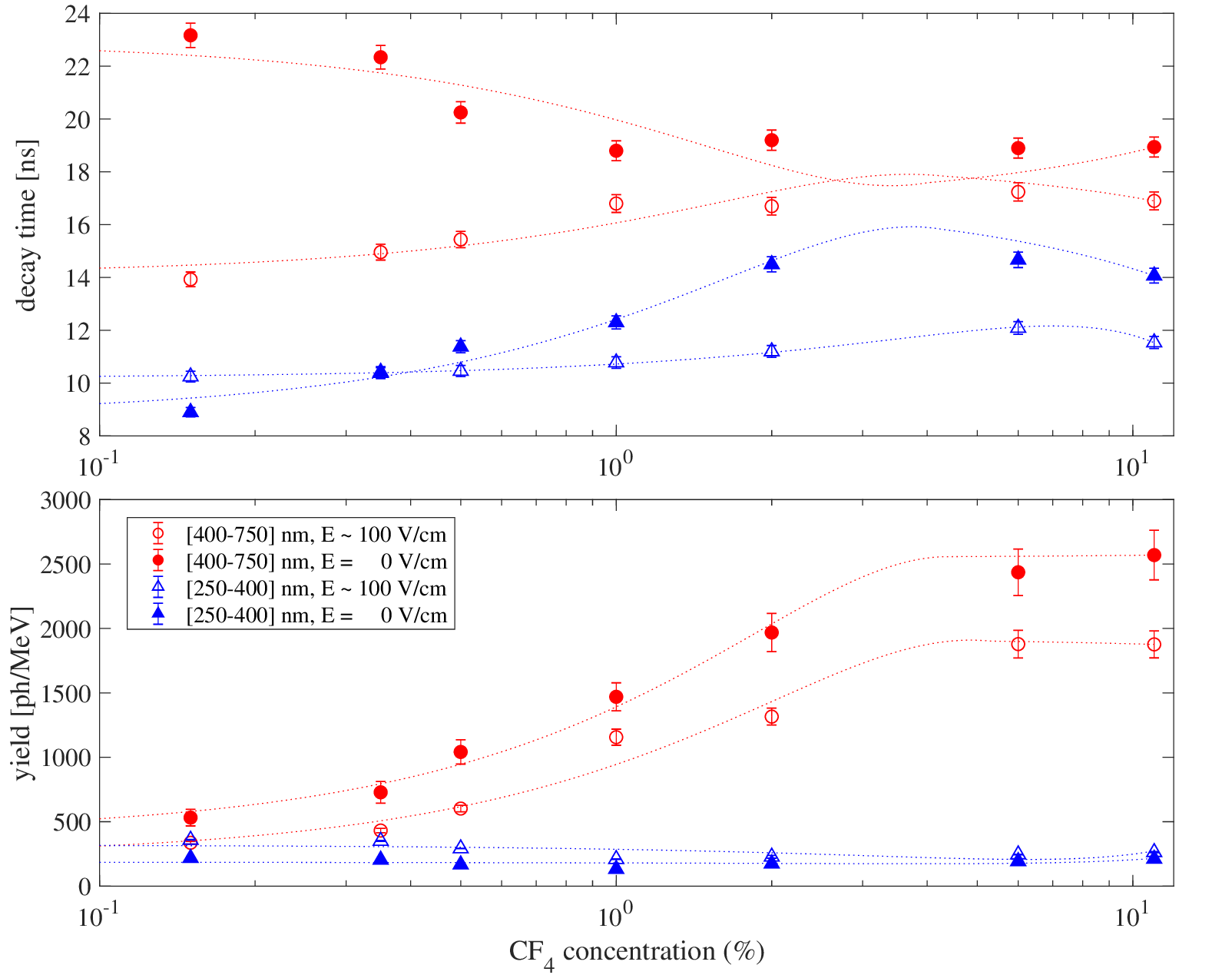}
    \caption{Compilation of the main scintillation characteristics of Ar/CF$_4$ mixtures under $\alpha$ particles from $^{241}$Am and at 9.5~bar, as a function of CF$_4$ concentration. Red symbols are used for measurements in the visible band and blue for the UV one. Solid symbols correspond to low-field measurements and open ones indicate high field. Top: decay times. Bottom: scintillation yields in ph/MeV. Dashed lines are splines to guide the eye.}
\label{fig:VsConc}
\end{figure}

Fig. \ref{fig:VsConc} summarizes the main results of this work, providing the time constants and yields for the UV and visible bands at low and high electric fields. The behaviour observed for the yields is similar to the one reported in \cite{Amedo2023} under X-rays, characterized by a sharp increase of the scintillation in the visible band up to an inversion point at around 10\% CF$_4$. The presence of the $3^{\textnormal{rd}}$ continuum emission, on the other hand, leads to a compensatory effect on the UV yields: the minimum of scintillation is observed at around 1\% CF$_4$ as the $3^{\textnormal{rd}}$ continuum is suppressed, followed by an increase with CF$_4$ concentration as the population of CF$_4^{+,*}$ ions increases. In the range explored, the UV yields change by less than a factor $\times 2$.

\begin{figure}[h!!!]
    \centering
    \includegraphics[scale=0.55]{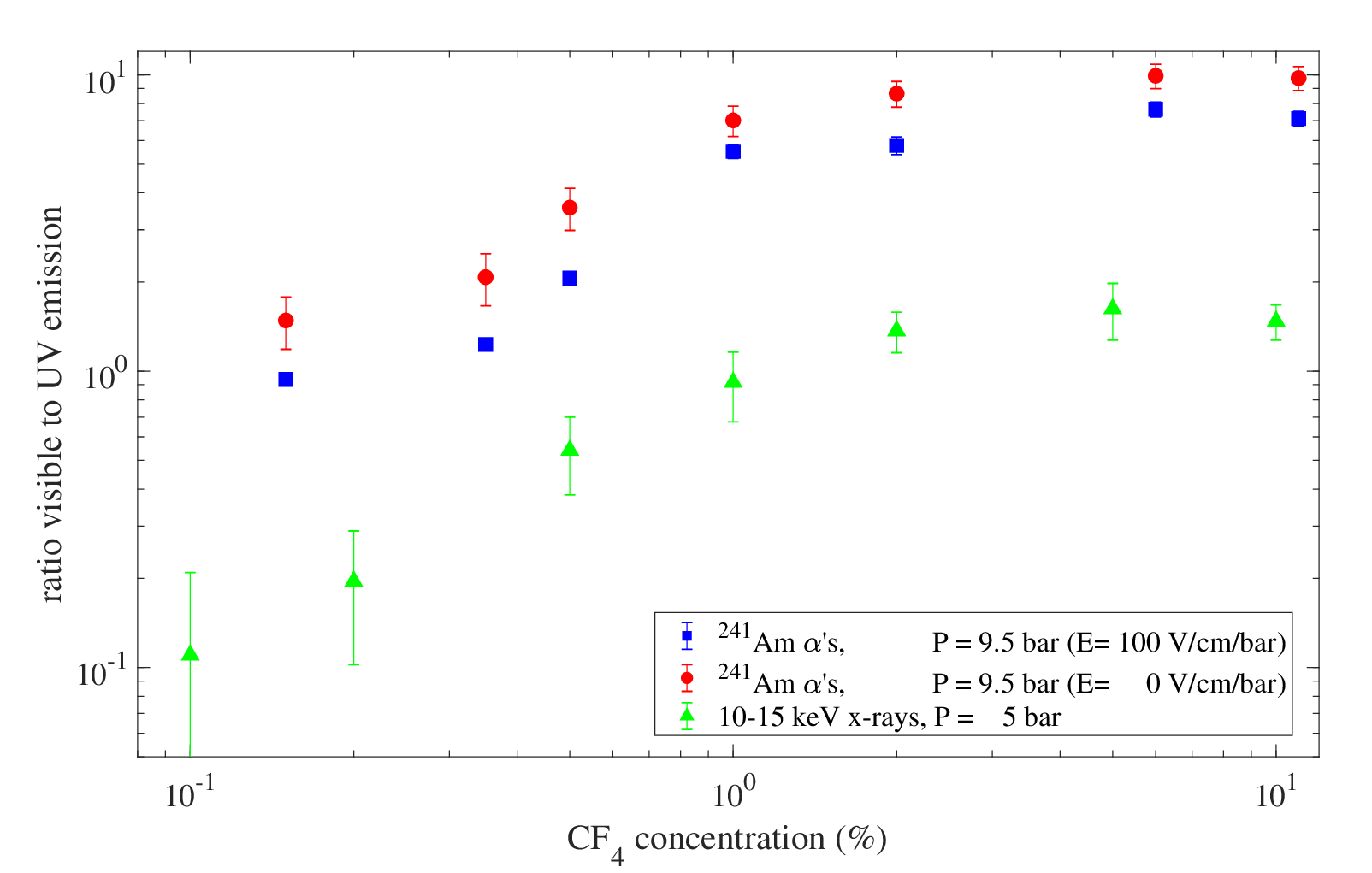}
    \caption{Ratio between scintillation in the visible and uv bands (vis/uv), defined in the range [400-750]~nm and [250-400]~nm. Blue (high field) and red (low field) markers show the results in this work for $\alpha$ particles (9.5~bar), while green markers show the result from previous measurements under X-rays (5~bar). The `ratio of ratios', $\mathcal{R}_{\alpha/x}$, stays at $4.6 \pm 0.8$(sta)$ \pm 1.4$(sys), compatible with earlier results.}
\label{fig:AlphasVSxrays}
\end{figure}

Despite qualitatively reproducing the concentration trends measured in X-ray data, the ratio of the yields in the visible and UV bands (vis/UV) that is observed under $\alpha$ particles in these measurements tops at nearly $\times 10$, much higher than the one observed for X-rays in \cite{Amedo2023}. A comparison is shown in Fig. \ref{fig:AlphasVSxrays}, displaying the ratio:

\begin{equation}
\mathcal{R}_{\alpha/x} = \frac{\textnormal{vis/uv}|_{\alpha}}{\textnormal{vis/uv}|_{x}} = 4.6 \pm 0.8
\end{equation}
Upon including the systematic uncertainty on present measurements, the estimate is $\mathcal{R}_{\alpha/x} = 4.6 \pm 0.8(sta) \pm 1.4(sys)$. Despite the different operating pressure, this result would hint at compatibility with earlier values of $\mathcal{R}_{\alpha/x} = 2.8\pm 0.3$ for spectral measurements performed for pure CF$_4$ at 1~bar. In all, present measurements further support the evidence of $\alpha$ particles leading to substantially higher vis/uv scintillation ratios compared to X-rays, and make it compelling to address the scintillation spectra of these mixtures under minimum ionizing radiation in the future.

\section{Conclusions}

We have presented measurements of the scintillation characteristics of Ar/CF$_4$ mixtures under $\alpha$-particle irradiation, at a pressure of 9.5~bar. The results are of potential interest to next-generation TPCs for neutrino physics, in which time stamping through primary scintillation is a valuable asset, needed to properly assign the time of the interaction to the neutrino beam. Our results show that, upon the mere addition of 1\%CF$_4$ to argon, scintillation yields in the range 1400-1200 ph/MeV can be obtained, in the range of pressure-reduced electric fields of [40-100]~V/cm/bar. The emission takes place predominantly in the [400-750]~nm band, easily accessible to silicon-based devices. The decay constant of the scintillation is much shorter than that of noble gases, in the range of $\tau = 14$-$17$~ns, thus showing good prospects for timing. Our measurements confirm earlier results obtained under X-rays, in regard to the gaseous wavelength-shifting capabilities of CF$_4$ when added to argon. The discrepancy between the intensity in the visible and UV bands as observed for $\alpha$'s and X-rays, $\mathcal{R}_{\alpha/x} = 4.6 \pm 0.8(sta) \pm 1.4(sys)$, is compatible with previous observations. It makes a compelling case for the measurement of scintillation under minimum ionizing particles and $\alpha$'s in the same conditions. 

\acknowledgments

This research has received financial support from the European Union’s Horizon 2020 Research and Innovation programme under GA no.\ 101004761, from Xunta de Galicia (Centro singular de investigación de Galicia, accreditation 2019-2022), and by the “María de Maeztu” Units of Excellence program MDM-2016-0692. DGD was supported by the Ram\'on y Cajal program (Spain) under contract number RYC-2015-18820. This research was also partly funded by the Spanish Ministry (‘Proyectos de Generación de Conocimiento’, PID2021-125028OB-C21).

Special thanks must be given to Alan Bross, Carlos Escobar and Adam Para (Fermilab) for encouragement and many insightful discussions.


\bibliographystyle{jhep}
\bibliography{bibliography_CAPS.bib}

\providecommand{\href}[2]{#2}\begingroup\raggedright\begin{thebibliography}{10}

\bibitem{Battat2014}
J.B.~Battat, C.~Deaconu, G.~Druitt, R.~Eggleston, P.~Fisher, P.~Giampa et~al., \emph{{The Dark Matter Time Projection Chamber 4Shooter directional dark matter detector: Calibration in a surface laboratory}}, \href{https://doi.org/10.1016/j.nima.2014.04.010}{\emph{Nuclear Instruments and Methods in Physics Research, Section A: Accelerators, Spectrometers, Detectors and Associated Equipment} {\bfseries 755} (2014) 6}.

\bibitem{ARIADNE_CF4}
A.~Roberts, P.~Svihra, A.~Al-Refaie, H.~Graafsma, J.~Küpper, K.~Majumdar et~al., \emph{{First demonstration of 3D optical readout of a TPC using a single photon sensitive Timepix3 based camera}}, \href{https://doi.org/10.1088/1748-0221/14/06/P06001}{\emph{Journal of Instrumentation} {\bfseries 14} (2019) }.

\bibitem{Fraga:2002uc}
F.A.F.~Fraga, L.M.S.~Margato, S.T.~Fetal, M.M.F.R.~Fraga, R.~Ferreira-Marques, A.J.P.L.~Policarpo et~al., \emph{{CCD readout of GEM-based neutron detectors}}, \href{https://doi.org/10.1016/S0168-9002(01)01829-0}{\emph{Nucl. Instrum. Meth. A} {\bfseries 478} (2002) 357}.

\bibitem{Takahashi2011}
M.~Takahashi, S.~Kabuki, K.~Hattori, N.~Higashi, S.~Iwaki, H.~Kubo et~al., \emph{{Development of an Electron-Tracking Compton Camera using CF4 gas at high pressure for improved detection efficiency}}, \href{https://doi.org/10.1016/j.nima.2010.06.305}{\emph{Nuclear Instruments and Methods in Physics Research, Section A: Accelerators, Spectrometers, Detectors and Associated Equipment} {\bfseries 628} (2011) 150}.

\bibitem{RICH_LHCb}
R.~Forty, \emph{{RICH pattern recognition for LHCb}}, \href{https://doi.org/10.1016/S0168-9002(99)00310-1}{\emph{Nuclear Instruments and Methods in Physics Research Section A: Accelerators, Spectrometers, Detectors and Associated Equipment} {\bfseries 433} (1999) 257}.

\bibitem{Fraga_GEM}
M.~Fraga, F.~Fraga, S.~Fetal, L.~Margato, R.~Marques and A.~Policarpo, \emph{{The GEM scintillation in He–CF4, Ar–CF4, Ar–TEA and Xe–TEA mixtures}}, \href{https://doi.org/10.1016/S0168-9002(03)00758-7}{\emph{Nuclear Instruments and Methods in Physics Research Section A: Accelerators, Spectrometers, Detectors and Associated Equipment} {\bfseries 504} (2003) 88}.

\bibitem{Pansky1995}
A.~Pansky, A.~Breskin, A.~Buzulutskov, R.~Chechik, V.~Elkind and J.~Va'vra, \emph{{The scintillation of CF4 and its relevance to detection science}}, \href{https://doi.org/10.1016/0168-9002(94)01064-1}{\emph{Nuclear Inst. and Methods in Physics Research, A} {\bfseries 354} (1995) 262}.

\bibitem{Morozov2010}
A.~Morozov, M.M.~Fraga, L.~Pereira, L.M.~Margato, S.T.~Fetal, B.~Guerard et~al., \emph{{Photon yield for ultraviolet and visible emission from CF4 excited with $\alpha$-particles}}, \href{https://doi.org/10.1016/j.nimb.2010.01.012}{\emph{Nuclear Instruments and Methods in Physics Research, Section B: Beam Interactions with Materials and Atoms} {\bfseries 268} (2010) 1456}.

\bibitem{Azmoun:2010zz}
B.~Azmoun, A.~Caccavano, M.~Rumore, J.~Sinsheimer, N.~Smirnov, S.~Stoll et~al., \emph{{A measurement of the scintillation light yield in CF(4) using a photosensitive GEM detector}}, \href{https://doi.org/10.1109/TNS.2010.2052632}{\emph{IEEE Trans. Nucl. Sci.} {\bfseries 57} (2010) 2376}.

\bibitem{Lehaut2015}
G.~Lehaut, S.~Salvador, J.M.~Fontbonne, F.R.~Lecolley, J.~Perronnel and C.~Vandamme, \emph{{Scintillation properties of N2 and CF4 and performances of a scintillating ionization chamber}}, \href{https://doi.org/10.1016/j.nima.2015.05.050}{\emph{Nuclear Instruments and Methods in Physics Research, Section A: Accelerators, Spectrometers, Detectors and Associated Equipment} {\bfseries 797} (2015) 57}.

\bibitem{Saito_Ws_NobleGases_2002}
K.~Saito, H.~Tawara, T.~Sanami, E.~Shibamura and S.~Sasaki, \emph{{Absolute number of scintillation photons emitted by alpha particles in rare gases}}, \href{https://doi.org/10.1109/TNS.2002.801700}{\emph{IEEE Transactions on Nuclear Science} {\bfseries 49 I} (2002) 1674}.

\bibitem{Morozov2011}
A.~Morozov, M.~Fraga, L.~Pereira, L.~Margato, S.~Fetal, B.~Guerard et~al., \emph{{Effect of the electric field on the primary scintillation from CF4}}, \href{https://doi.org/10.1016/j.nima.2010.07.001}{\emph{Nuclear Instruments and Methods in Physics Research Section A: Accelerators, Spectrometers, Detectors and Associated Equipment} {\bfseries 628} (2011) 360}.

\bibitem{xenon_paper}
S.~Leardini, E.S.~García, P.~Amedo, A.~Saa-Hernández, D.~González-Díaz, R.~Santorelli et~al., \emph{{Time and band-resolved scintillation in time projection chambers based on gaseous xenon}}, \href{https://doi.org/10.1140/epjc/s10052-022-10385-y}{\emph{The European Physical Journal C} {\bfseries 82} (2022) 425}.

\bibitem{Santorelli2021}
R.~Santorelli, E.S.~Garcia, P.G.~Abia, D.~González-Díaz, R.L.~Manzano, J.J.~Morales et~al., \emph{{Spectroscopic analysis of the gaseous argon scintillation with a wavelength sensitive particle detector}}, \href{https://doi.org/10.1140/epjc/s10052-021-09375-3}{\emph{European Physical Journal C} {\bfseries 81} (2021) 622}.

\bibitem{Spectral_Atlas_2013}
H.~Keller-Rudek, G.K.~Moortgat, R.~Sander and R.~Sörensen, \emph{{The MPI-Mainz UV/VIS Spectral Atlas of Gaseous Molecules of Atmospheric Interest}}, \href{https://doi.org/10.5194/essd-5-365-2013}{\emph{Earth System Science Data} {\bfseries 5} (2013) 365}.

\bibitem{Amedo2023}
P.~Amedo, D.~González-Díaz, F.M.~Brunbauer, D.J.~Fernández-Posada, E.~Oliveri and L.~Ropelewski, \emph{{Observation of strong wavelength-shifting in the argon-tetrafluoromethane system}}, \href{https://doi.org/10.3389/fdest.2023.1282854}{\emph{Frontiers in Detector Science and Technology} {\bfseries 1} (2023) }.

\bibitem{Christophorou2004}
L.G.~Christophorou and J.K.~Olthoff, \emph{{Fundamental Electron Interactions with Plasma Processing Gases}}, Springer US (2004), \href{https://doi.org/10.1007/978-1-4419-8971-0}{10.1007/978-1-4419-8971-0}.

\bibitem{Amedo_tracks_2024}
P.~Amedo, R.~Hafeji, A.~Roberts, A.~Lowe, S.~Ravinthiran, S.~Leardini et~al., \emph{{Scintillation of Ar/CF4 mixtures: glass-THGEM characterization with 1\% CF4 at 1–1.5 bar}}, \href{https://doi.org/10.1088/1748-0221/19/05/C05001}{\emph{Journal of Instrumentation} {\bfseries 19} (2024) C05001}.

\bibitem{DiegoReview2018}
D.~González-Díaz, F.~Monrabal and S.~Murphy, \emph{{Gaseous and dual-phase time projection chambers for imaging rare processes}}, \href{https://doi.org/10.1016/j.nima.2017.09.024}{\emph{Nuclear Instruments and Methods in Physics Research, Section A: Accelerators, Spectrometers, Detectors and Associated Equipment} {\bfseries 878} (2018) 200}.

\bibitem{Sain2010_Penning}
{\"O}.~Şahin, I.~Tapan, N.~Özmutlu and R.~Veenhof, \emph{{Penning transfer in argon-based gas mixtures}}, \href{https://doi.org/10.1088/1748-0221/5/05/P05002}{\emph{Journal of Instrumentation} {\bfseries 5} (2010) P05002}.

\bibitem{Sahin2016_Penning}
{\"O}.~Şahin, T.Z.~Kowalski, T.~Murakami, K.~Niemi, T.~Gans and R.~Veenhof, \emph{{Systematic gas gain measurements and Penning energy transfer rates in Ne-CO2 mixtures}}, \href{https://doi.org/10.1088/1748-0221/11/01/P01003}{\emph{Journal of Instrumentation} {\bfseries 11} (2016) P01003}.

\bibitem{Forster_1959}
T.~Főrster, \emph{{10th Spiers Memorial Lecture. Transfer mechanisms of electronic excitation}}, \href{https://doi.org/10.1039/DF9592700007}{\emph{Discuss. Faraday Soc.} {\bfseries 27} (1959) 7}.

\bibitem{Benson2018}
C.~Benson, G.O.~Gann and V.~Gehman, \emph{{Measurements of the intrinsic quantum efficiency and absorption length of tetraphenyl butadiene thin films in the vacuum ultraviolet regime}}, \href{https://doi.org/10.1140/epjc/s10052-018-5807-z}{\emph{European Physical Journal C} {\bfseries 78} (2018) }.

\bibitem{Kuzniak_2021_FATGEMS}
M.~Kuźniak, D.~González-Díaz, P.~Amedo, C.D.~Azevedo, D.J.~Fernández-Posada, M.~Kuźwa et~al., \emph{{Development of very-thick transparent GEMs with wavelength-shifting capability for noble element TPCs}}, \href{https://doi.org/10.1140/EPJC/S10052-021-09316-0}{\emph{The European Physical Journal C 2021 81:7} {\bfseries 81} (2021) 1}.

\bibitem{DUNE_Xe_Doping_2024}
A.A.~Abud, B.~Abi, R.~Acciarri, M.~Acero, M.~Adames, G.~Adamov et~al., \emph{{Doping liquid argon with xenon in ProtoDUNE Single-Phase: effects on scintillation light}}, \href{https://doi.org/10.1088/1748-0221/19/08/P08005}{\emph{Journal of Instrumentation} {\bfseries 19} (2024) P08005}.

\bibitem{Segreto_TPB_LAr_2015}
E.~Segreto, \emph{{Evidence of delayed light emission of tetraphenyl-butadiene excited by liquid-argon scintillation light}}, \href{https://doi.org/10.1103/PhysRevC.91.035503}{\emph{Physical Review C} {\bfseries 91} (2015) 035503}.

\bibitem{CYGNUS}
S.E.~Vahsen, C.A.J.~O'Hare, W.A.~Lynch, N.J.C.~Spooner, E.~Baracchini, P.~Barbeau et~al., \emph{{CYGNUS: Feasibility of a nuclear recoil observatory with directional sensitivity to dark matter and neutrinos}}, \href{https://doi.org/https://doi.org/10.48550/arXiv.2008.12587}{\emph{arXiv:2008.12587} (2020) }.

\bibitem{Migdal_2023}
H.M.~Araújo, S.N.~Balashov, J.E.B.F.M.~Brunbauer, C.~Cazzaniga, C.D.~Frost, F.~Garcia et~al., \emph{{The MIGDAL experiment: Measuring a rare atomic process to aid the search for dark matter}}, \href{https://doi.org/10.1016/j.astropartphys.2023.102853}{\emph{Astroparticle Physics} {\bfseries 151} (2022) 102853}.

\bibitem{NDGAR_Snowmass2021}
A.A.~Abud, B.~Abi, R.~Acciarri, M.A.~Acero, M.R.~Adames, G.~Adamov et~al., \emph{{A Gaseous Argon-Based Near Detector to Enhance the Physics Capabilities of DUNE}}, \href{https://doi.org/https://doi.org/10.48550/arXiv.2203.06281}{\emph{arXiv:2203.06281} (2022) }.

\bibitem{Cortesi2023}
M.~Cortesi, H.~Sims, J.~Pereira, Y.~Ayyad, P.~Majewski and I.~Katsioulas, \emph{{Secondary scintillation properties of multi-layer THGEMs operated in low-pressure CF 4 and Ar/5\%Xe}}, \href{https://doi.org/10.1088/1748-0221/18/08/P08005}{\emph{Journal of Instrumentation} {\bfseries 18} (2023) P08005}.

\bibitem{DUNE_Phase_II}
A.A.~Abud, B.~Abi, R.~Acciarri, M.A.~Acero, M.R.~Adames, G.~Adamov et~al., \emph{{DUNE Phase II: Scientific Opportunities, Detector Concepts, Technological Solutions}}, \href{https://doi.org/https://doi.org/10.48550/arXiv.2408.12725}{\emph{arXiv:2408.12725} (2024) }.

\bibitem{COHERENT_Snowmass_2021}
J.~Asaadi, P.S.~Barbeau, B.~Bodur, A.~Bross, E.~Conley, Y.~Efremenko et~al., \emph{{Physics Opportunities in the ORNL Spallation Neutron Source Second Target Station Era}}, \href{https://doi.org/https://doi.org/10.48550/arXiv.2209.02883}{\emph{arXiv:2209.02883} (2022) }.

\bibitem{Ar_CF4_Timing_2024}
A.~Saá-Hernández, D.~González-Díaz, J.~Martín-Albo, M.~Tuzi, P.~Amedo, J.~Baldonedo et~al., \emph{{On the determination of the interaction time of GeV neutrinos in large argon gas TPCs}}, \href{https://doi.org/https://doi.org/10.48550/arXiv.2401.09920}{\emph{arXiv:2401.09920} (2024) }.

\bibitem{Margato}
L.M.~Margato, A.~Morozov, L.~Pereira, M.M.~Fraga and F.A.~Fraga, \emph{{Effect of the gas contamination on CF 4 primary and secondary scintillation}}, \href{https://doi.org/10.1016/j.nima.2011.10.033}{\emph{Nuclear Instruments and Methods in Physics Research, Section A: Accelerators, Spectrometers, Detectors and Associated Equipment} {\bfseries 695} (2012) 425}.

\end{thebibliography}\endgroup

\end{document}